\documentclass[prd,superscriptaddress,amsfonts,amssymb,amsmath,showpacs,twocolumn]{revtex4-2}
\usepackage{bm}
\usepackage{amsfonts}
\usepackage{latexsym}
\usepackage{graphicx}
\usepackage{amsmath}
\usepackage{palatino}
\usepackage{mathpazo}
\usepackage{textcomp}
\usepackage{float}
\usepackage{booktabs}
\usepackage{dcolumn}
\usepackage{booktabs}
\usepackage{multirow}
\usepackage{hyperref}
\hypersetup{colorlinks,citecolor=blue}
\usepackage{amsmath}
\usepackage{xcolor}
\usepackage{orcidlink}
\usepackage{epsfig}
\usepackage{caption}
\usepackage{subcaption}
\usepackage{commath}
\def\jnl@style{\it}
\def\aaref@jnl#1{{\jnl@style#1}}

\def\aaref@jnl#1{{\jnl@style#1}}

\def\aj{\aaref@jnl{AJ}}                   % Astronomical Journal
\def\apj{\aaref@jnl{ApJ}}                 % Astrophysical Journal
\def\apjl{\aaref@jnl{ApJ}}                % Astrophysical Journal, Letters
\def\apjs{\aaref@jnl{ApJS}}               % Astrophysical Journal, Supplement
\def\apss{\aaref@jnl{Ap\&SS}}             % Astrophysics and Space Science
\def\aap{\aaref@jnl{A\&A}}                % Astronomy and Astrophysics
\def\aapr{\aaref@jnl{A\&A~Rev.}}          % Astronomy and Astrophysics Reviews
\def\aaps{\aaref@jnl{A\&AS}}              % Astronomy and Astrophysics, Supplement
\def\mnras{\aaref@jnl{Mon.~Not.~Roy.~Astron.~Soc.}}             % Monthly Notices of the RAS
\def\prd{\aaref@jnl{Phys.~Rev.~D}}        % Physical Review D
\def\prc{\aaref@jnl{Phys.~Rev.~C}}  % Physical Review C
\def\prl{\aaref@jnl{Phys.~Rev.~Lett.}}    % Physical Review Letters
\def\qjras{\aaref@jnl{QJRAS}}             % Quarterly Journal of the RAS
\def\skytel{\aaref@jnl{S\&T}}             % Sky and Telescope
\def\ssr{\aaref@jnl{Space~Sci.~Rev.}}     % Space Science Reviews
\def\zap{\aaref@jnl{ZAp}}                 % Zeitschrift fuer Astrophysik
\def\nat{\aaref@jnl{Nature}}              % Nature
\def\aplett{\aaref@jnl{Astrophys.~Lett.}} % Astrophysics Letters
\def\apspr{\aaref@jnl{Astrophys.~Space~Phys.~Res.}} % Astrophysics Space Physics Research
\def\physrep{\aaref@jnl{Phys.~Rep.}}      % Physics Reports
\def\physscr{\aaref@jnl{Phys.~Scr}}       % Physica Scripta
\def\commat{\aaref@jnl{Comm.~Math.~Phys.}}              % Communications in Mathematical Physics
\def\science{\aaref@jnl{Science}}               % Science
\def\cqg{\aaref@jnl{Classical Quant.~Grav.}}            % Classical and Quantum Gravity
\def\jpcs{\aaref@jnl{JPCS}}                                     % Journal of Physics Conference Series
\def\ijmpd{\aaref@jnl{Int.~J.~Mod.~Phys.~D}}                    % International Journal of Modern Physics D
\def\grg{\aaref@jnl{Gen.~Relat.~Gravit.}}               % General Relativity and Gravitation
\def\rpp{\aaref@jnl{Rep.~Prog.~Phys.}}          % Reports on Progress in Physics
\def\npa{\aaref@jnl{Nucl.~Phys.~A}}        % Nuclear Physics A
\def\lrr{\aaref@jnl{Living Rev.~Rel.}}                   % Living reviews in relativity
\def\jcap{\aaref@jnl{J.~Cosmology Astropart.~Phys.}}    % Journal of cosmology and astroparticle physics
\def\rmp{\aaref@jnl{Rev.~Mod.~Phys.}}   %Reviews of modern physics
\def\epjc{\aaref@jnl{Eur.~Phys.~J.~C}}

\allowdisplaybreaks[1]
\begin{document}

\color{black}       %% For one column

\title{Modeling cosmic acceleration with a generalized varying deceleration parameter}

\author{M. Koussour\orcidlink{0000-0002-4188-0572}}
\email[Email: ]{pr.mouhssine@gmail.com}
\affiliation{Quantum Physics and Magnetism Team, LPMC, Faculty of Science Ben
M'sik,\\
Casablanca Hassan II University,
Morocco.} 

\author{N. Myrzakulov\orcidlink{0000-0001-8691-9939}}
\email[Email: ]{nmyrzakulov@gmail.com}
\affiliation{L. N. Gumilyov Eurasian National University, Astana 010008,
Kazakhstan.}
\affiliation{Ratbay Myrzakulov Eurasian International Centre for Theoretical
Physics, Astana 010009, Kazakhstan.}

\author{Alnadhief H. A. Alfedeel\orcidlink{0000-0002-8036-268X}}%
\email[Email: ]{aaalnadhief@imamu.edu.sa}
\affiliation{Department of Mathematics and Statistics, Imam Mohammad Ibn Saud Islamic University (IMSIU),\\
Riyadh 13318, Saudi Arabia.}
\affiliation{Department of Physics, Faculty of Science, University of Khartoum, P.O. Box 321, Khartoum 11115, Sudan.}
\affiliation{Centre for Space Research, North-West University, Potchefstroom 2520, South Africa.}

\author{F. Awad}%
\email[Email: ]{awad.fga@gmail.com}
\affiliation{Department of Mathematics, Omdurman Islamic Univerity, Khartoum, Sudan.}
\affiliation{Alneelain Center for Mathematical Sceinces, Alaneelain University, Kartoum, Sudan.}

\author{M. Bennai\orcidlink{0000-0002-7364-5171}}
\email[Email: ]{mdbennai@yahoo.fr}
\affiliation{Quantum Physics and Magnetism Team, LPMC, Faculty of Science Ben
M'sik,\\
Casablanca Hassan II University,
Morocco.} 
\affiliation{Lab of High Energy Physics, Modeling and Simulations, Faculty of
Science,\\
University Mohammed V-Agdal, Rabat, Morocco.}

%{Universidade Federal do ABC (UFABC) - Centro de Ci\^encias Naturais e Humanas (CCNH) - Avenida dos Estados 5001, 09210-580, Santo Andr\’e, SP, Brazil}
%\affiliation{Universidade de S\~ao Paulo (USP), Instituto de Astronomia, Geof\'isica e Ci\^encias Atmosf\'ericas (IAG), Rua do Mat\~ao 1226, Cidade Universit\'aria, 05508-090 S\~ao Paulo, SP, Brazil}
%\affiliation{Instituto Tecnol\'ogico de Aeron\'autica (ITA), Departamento de F\'isica, Centro T\'ecnico Aeroespacial, 12228-900 S\~ao Jos\'e dos Campos, S\~ao Paulo, Brazil}

%%%%%%%%%%%%%%%%%%%%%%%%%%%%%%%%%%%%  DATE  %%%%%%%%%%%%%%%%%%%%%%%%%%%%%%%%%%%%
\date{\today}

\begin{abstract}
Understanding the accelerating expansion of the Universe remains a fundamental challenge in modern cosmology. In this paper, we investigate a cosmological model parametrized by a generalized variable deceleration parameter to elucidate the dynamics driving cosmic acceleration. By employing constraints from the latest observational datasets, including Cosmic Chronometers (CC), Type Ia Supernovae (SNe), and Baryon Acoustic Oscillations (BAO), we assess the compatibility of the model with observational data. The chosen parametrization aligns with thermodynamic constraints on the deceleration parameter, further validating its reliability. Further, we estimate the present value of the Hubble parameter, transition redshift, deceleration parameter, and EoS parameter, which align with observational data. Lastly, our stability analysis confirms the model's stability against small perturbations.

\end{abstract}

\maketitle

\section{Introduction}
\label{sec1}

The observation of cosmic expansion accelerating is regarded as a significant breakthrough in modern cosmology \cite{Riess,Perlmutter,D.J.,W.J.,R.R.,Z.Y.,T.Koivisto,S.F.}. This acceleration is caused by an unknown force, called DE (Dark Energy), which is believed to permeate all of space and push the Universe's expansion with a repulsive force. The nature of DE remains a mystery and is an active area of research \cite{Frieman}. Modified gravity theories, such as the so-called geometrical modification to GR (General Relativity) theory, have been proposed to explain the late-time cosmic acceleration \cite{R1,T1,T2,Q1,Q2,Q3}.  In the standard $\Lambda$CDM ($\Lambda$-Cold Dark Matter) cosmological model, the total mass-energy of the Universe is composed of 4.9\% BM (Baryonic Matter), 26.8\% DM (Dark Matter), and 68.3\% of DE. DM is an unknown form of matter that only interacts gravitationally and does not emit or absorb light \cite{DM1}. Its existence is supported by astrophysical observations, although its nature remains unknown \cite{DM2}. The non-baryonic matter is believed to make up the majority of DM \cite{DM3}.

There exists a multitude of models aimed at elucidating the perplexing nature of DE, which is responsible for the observed acceleration of cosmic expansion. These models include the quintessence \cite{RP}, phantom \cite{M.S.,M.S.-2}, K-essence \cite{T.C.,C.A.}, Chaplygin gas \cite{M.C.,A.Y.}, tachyon \cite{T.P.}, and holographic DE model. The simplest and most widely used model is the cosmological constant model, which is consistent with observations \cite{Peb}. However, the cosmological constant $\Lambda$ encounters two challenging predicaments, namely fine-tuning and the cosmic coincidence problem. According to quantum field theory, the calculated value of vacuum energy \cite{S.W.} exceeds its observed value by a staggering $123$ orders of magnitude, where the observed value is on the order of $10^{-47} GeV^4 $ \cite{Riess, Perlmutter}.

Despite the existence of various theoretical approaches, the exact explanation for cosmic acceleration remains elusive. One prominent model employed to study late-time cosmic acceleration is reconstruction, which employs an inverse methodology to identify a suitable cosmological model. Reconstruction can be classified into two categories: parametric reconstruction and non-parametric reconstruction, both of which involve incorporating observational data directly into the construction of the model \cite{Mukherjee}. Parametric reconstruction, also known as the model-dependent approach, involves estimating model parameters based on diverse observational data. The underlying concept is to assume a specific evolutionary scenario and then determine the properties of the matter sector or exotic component responsible for the observed acceleration. Numerous researchers have employed this method to identify viable solutions and constraints \cite{Mamon1,Mamon2,Das}. The kinematic approach to studying cosmic evolution is agnostic to any specific gravity theory, allowing for an independent analysis \cite{Riess1}. Kinematic quantities, such as the deceleration parameter, the Hubble parameter, and the jerk parameter, are used to characterize the dynamics of the Universe. By reconstructing these kinematic quantities using observational data, we can gain insights into the nature of cosmic evolution without making any assumptions about DE or specific gravity theories. This approach provides a model-independent framework for understanding the behavior of the Universe.

The parameterization of the deceleration parameter $q$ is a crucial aspect in understanding the evolution of the Universe's expanding rate. It provides insight into the possible future of the Universe, including scenarios where the Universe may experience accelerated expansion or eventually collapse. To accommodate both scenarios of the Universe, the deceleration parameter must exhibit a distinct change. Specifically, there needs to be a transition from a decelerating phase ($q > 0$) to a late-time accelerating phase ($q < 0$). This transition is essential for explaining the formation of cosmic structures and aligning with the observed acceleration of the Universe. By capturing this signature flip in the deceleration parameter, we can reconcile both aspects of cosmic evolution. Many studies have used different parametric forms of deceleration parameters, while others have explored non-parametric forms. Furthermore, the choice of the parametric form of the deceleration parameter can affect the accuracy of cosmological measurements, such as the estimation of the Hubble constant and the age of the Universe. This underscores the importance of choosing a suitable parametric form and the need for further research in this area. The literature has extensively discussed these methods to address various cosmological issues, such as the initial singularity problem, the problem of all-time decelerating expansion, the horizon problem, Hubble tension, and more \cite{Singh,Sanchez,Dinda}. Through these investigations, researchers have gained valuable insights into the fundamental properties of the Universe and how they have evolved over time. 

In the literature, several commonly used parametrizations of the deceleration parameter have been proposed and employed in cosmological studies. The linear parametrization assumes a linear relationship between the deceleration parameter and redshift. It is expressed as $q(z) = q_0 + q_1z$, where $q_0$ and $q_1$ are the model parameters. This parametrization captures a simple transition from a decelerating phase to an accelerating phase and has been widely used in various cosmological analyses \cite{Riess1,Akarsu}. The CPL (Chevallier-Polarski-Linder) parametrization is a widely used parametrization that was proposed to describe the equation of state (EoS) of DE. It can also be applied to the deceleration parameter by assuming a specific functional form i.e. $q(z) = q_{0} + q_{1} z (1 + z)^{-1}$. The CPL parametrization captures a variety of dynamics, including both phantom and quintessence behavior, and is compatible with several theoretical models \cite{Santos}. Recently, Del Campo et al. \cite{Campo} introduces three parametrizations of the deceleration parameter based on thermodynamic principles. The study explores the thermodynamic properties and implications of these parametrizations, considering the dynamical evolution of the Universe. The authors provide mathematical expressions and analyze the cosmological consequences of each parametrization, comparing them with existing models. In addition, various parameterizations have been proposed in the literature including forms such as $q(z) = q_{1} + q_{2}z(1+z)^{-2}$, $q(z) = 1/2 + q_{1}(1+z)^{-2}$, $q(z) = 1/2 + (q_{1}z + q_{2})(1+z)^{-2}$, and more complex expressions \cite{Nair,Gong,Xu}. In contrast to many existing parametrizations of the deceleration parameter that have limitations in their validity range, this present paper introduces a generalized variable deceleration parameter that addresses these limitations. The proposed parametrization, consisting of three free parameters, is applicable from the matter-dominated epoch ($z \gg 1$) onwards, extending up to $z = -1$. This choice is motivated by practical and theoretical considerations, ensuring independence from specific cosmological models \cite{CapozzielloDP}. Notably, the parametrization satisfies asymptotic conditions: $q(z \gg 1) = 1/2$, $q(z = -1) = -1$, and an additional condition, $dq/dz > 0$, which holds at least in the limit $q \rightarrow -1$. This generalized approach provides a more comprehensive and consistent description of the deceleration parameter throughout a wide redshift range. 

The structure of this paper is outlined as follows. In Sec. \ref{sec2}, we present the fundamental theoretical framework for the scalar field DE model within the context of a spatially flat FLRW (Friedmann-Lemaître-Robertson-Walker) Universe. We adopt a specific form of the deceleration parameter, denoted as $q(z)$, to solve the field equations. Furthermore, we employ this parametrization to reconstruct the EoS associated with the scalar field. Sec. \ref{sec3} of the paper focuses on the inclusion of observational datasets, namely Cosmic Chronometers (CC), Type Ia Supernovae (SNe), and Baryon Acoustic Oscillations (BAO), for our analysis. We carefully consider these datasets and utilize them to impose constraints on the different model parameters that we have reconstructed. The details of the data analysis and the methodology employed are thoroughly discussed in this section. In Sec. \ref{sec4}, we present the main results derived from our analysis. We discuss the implications of the constrained model parameters and examine their consistency with the observational data. In Sec. \ref{sec5}, we delve into the stability analysis of the proposed model. We examine the behavior of the system under small perturbations and assess its overall stability. In the final section \ref{sec6}, we present the conclusions drawn from our study based on the obtained results. 

In this paper, we have followed the convention of adopting the natural units where the constants $8\pi G$ and $c$ are set to unity. 

\section{Overview of the model and foundational equations}
\label{sec2}

Scalar fields with positive potentials are often considered as possible
candidates for DE \cite{Copeland,Tsujikawa}. These fields have already been used in the
context of inflationary models, which seek to elucidate the origins of the large-scale structure observed in the Universe \cite{Turner,Linde}. In these models, the Universe
undergoes a rapid quasi-exponential expansion, during which small quantum
fluctuations are stretched to cosmological scales, providing a mechanism for the process of structure formation, encompassing the emergence of cosmic entities like galaxies and galaxy clusters. In the standard inflationary scenario \cite{Chervon}, the Universe is
dominated by a real scalar field $\phi$ that is homogeneously distributed in
space and has a potential function $V(\phi)$. The field responsible for
driving this expansion is known as the inflaton and the corresponding action
is described by 
\begin{equation}
S = \int\sqrt{-g} d^{4}x{\left(\frac{R}{2} - \frac{1}{2}g^{\mu \nu }\partial
_{\mu }\phi \partial _{\nu }\phi -V\left( \phi \right) \right)+S_{m}},
\label{Action}
\end{equation}
where $S_{m}$ represents the action of ordinary matter, $g$ represents the
determinant of the metric tensor $g_{\mu \nu }$, and $R$ represents the
Ricci (curvature) scalar. Thus, the Lagrangian density of a scalar field is 
\begin{equation}
L_{\phi}=- \frac{1}{2}g^{\mu \nu }\partial _{\mu }\phi \partial _{\nu }\phi
-V\left( \phi\right).
\end{equation}

The field equations can be obtained by varying Eq. (\ref{Action}) with
respect to $g_{\mu \nu }$, 
\begin{equation}
R_{_{\mu \nu }}-\frac{1}{2}g_{_{\mu \nu }}R=\left( T_{\mu \nu }^{(m)}+T_{\mu
\nu }^{(\phi )}\right) ,  \label{EFE}
\end{equation}%
where 
\begin{equation}
T_{\mu \nu }^{(m)}=\left( \rho_{m} +p_{m}\right) u_{\mu }u_{\nu }+p_{m}g_{\mu \nu },
\end{equation}%
and%
\begin{equation}
T_{\mu \nu }^{(\phi )}=\partial _{\mu }\phi \partial _{\nu }\phi -g_{\mu \nu
}\left( \frac{1}{2}g^{\rho \sigma }\partial _{\rho }\phi \partial _{\sigma
}\phi +V\left( \phi \right) \right) ,
\end{equation}%
are the energy- momentum tensor of ordinary matter and the scalar field,
respectively. Here, $u^{\mu }$ represents the 4-velocity of a comoving
observer, $\rho_{m} $ is the energy density of ordinary matter considered as a
perfect fluid and $p_{m}$ its pressure. In addition, the equation of motion for
the scalar field can be obtained by varying Eq. (\ref{Action}) with respect
to $\phi $,%
\begin{equation}
\Box \phi -V^{\prime }\left( \phi \right) =0.  \label{ME}
\end{equation}

Here, the D'Alembertian operator $\Box $ is defined by $\Box \equiv
\nabla ^{\mu }\nabla _{\mu }$, where $\nabla _{\mu }$ represents the
covariant derivative and $V^{\prime }\left( \phi \right) =\frac{\partial
V\left( \phi \right) }{\partial \phi }$ represents the derivative of the
potential $V\left( \phi \right) $ with respect to $\phi $. Also, one can
express the metric for the FLRW
spacetime as,%
\begin{equation}
ds^{2}=-dt^{2}+a^{2}\left( t\right) \left( \frac{dr^{2}}{1-kr^{2}}%
+r^{2}\left( d\theta ^{2}+\sin ^{2}\theta d\varphi ^{2}\right) \right) .
\end{equation}

Here, we have chosen pseudo-spherical coordinates ($r,\theta ,\varphi $), $%
k=-1,0,1$\ is the spatial curvature indicating an open, flat, or closed
Universe, respectively, and $a\left( t\right) $ is the scale factor of the
Universe. Hence, the field equations for a flat Universe with the presence
of a scalar field in the FLRW metric are given by \cite{Barrow1,Barrow2}
\begin{equation}
3H^{2}=\rho_{m} +\frac{1}{2}\overset{.}{\phi }^{2}+V\left( \phi \right) ,
\label{F1}
\end{equation}%
\begin{equation}
2\overset{.}{H}+3H^{2}=-p_{m}-\frac{1}{2}\overset{.}{\phi }^{2}+V\left( \phi
\right) ,  \label{F2}
\end{equation}%
while the motion equation (\ref{ME}), which has the form%
\begin{equation}
\overset{..}{\phi }+3H\overset{.}{\phi }+V^{\prime }\left( \phi \right) =0,
\end{equation}%
where $H=\frac{\overset{.}{a}}{a}$ is the Hubble parameter, describes the
expansion rate of the Universe.

By observing Eqs. (\ref{F1}) and (\ref{F2}), it can be noted that the scalar
field $\phi $ contributes to the energy content of the Universe in the form
of a perfect fluid, with pressure $p_{\phi }=\omega _{\phi }\rho _{\phi }$,
with energy density and pressure given by%
\begin{equation}
\rho _{\phi }=\frac{1}{2}\overset{.}{\phi }^{2}+V\left( \phi \right) ,
\end{equation}%
\begin{equation}
p_{\phi }=\frac{1}{2}\overset{.}{\phi }^{2}-V\left( \phi \right) ,
\end{equation}%
respectively. The EoS parameter of the scalar field is
then given by,%
\begin{equation}
\omega _{\phi }=\frac{\overset{.}{\phi }^{2}-2V\left( \phi \right) }{\overset%
{.}{\phi }^{2}+2V\left( \phi \right) }.  \label{Omega_phi}
\end{equation}

Eq. (\ref{Omega_phi}) suggests that scalar fields that evolve slowly at $%
\overset{.}{\phi }\rightarrow 0$, are difficult to distinguish from the
cosmological constant. Alternatively, the cosmological constant can be
viewed as a unique scenario of a constant scalar field. These models based
on scalar fields that aim to explain the cosmic acceleration in the
late-time Universe are referred to as quintessence models.

One way to describe the overall behavior of DE in the Universe is to use the total equation-of-state parameter, which can be defined as
\begin{equation}
    \omega=\frac{p}{\rho}=\frac{p_{\phi}+p_{m}}{\rho_{\phi}+\rho_{m}}.
\end{equation}

We assume $p_{m}=0$ for the dust Universe, the total EoS can be written as $\omega=\frac{p_{\phi}}{\rho_{\phi}+\rho_{m}}$. The conservation equations for the scalar field and matter can be written as follows:
\begin{equation}
\overset{.}{\rho }_{\phi}+3\left( \rho _{\phi}+p_{\phi}\right) H=0.
\label{cf}
\end{equation}
\begin{equation}
\overset{.}{\rho }_{m}+3\rho _{m}H=0.  \label{cm}
\end{equation}%

Therefore, the energy density associated with matter can be expressed as
\begin{equation}
\rho _{m}=\frac{\rho_{m0}}{a^3}=\rho _{m0}\left( 1+z\right) ^{3},
\label{rhom}
\end{equation}%
where $\rho_{m0}$ is an integration constant that signifies the energy density of matter at the present time.

Now, we can define the dimensionless density parameters corresponding to the scalar field and matter density as follows:
\begin{equation}
    \Omega_{\phi}=\frac{\rho_{\phi}}{3H^{2}},  \qquad \Omega_{m}=\frac{\rho_{m}}{3H^{2}}
\end{equation}

As we mentioned before, the study of cosmological models relies heavily on kinematic variables. The
deceleration parameter, for instance, characterizes the behavior of the
Universe, such as whether it is undergoing deceleration, acceleration, or a
transition phase. The EoS parameter $\omega $ provides insight into the
physical properties of the energy sources driving the evolution of the
Universe. In addition, to determine other cosmological parameters, such as
pressure, energy densities, EoS parameter, and potential function, an extra
equation is needed to complete the system of field equations i.e. Eqs. (\ref%
{F1}) and (\ref{F2}). This supplementary equation can be any functional form
of a cosmological parameter, such as the Hubble parameter, deceleration
parameter, and EoS parameter, providing necessary constraint equations \cite{Pacif}.
Here, we assume a generalized variable deceleration parameter of the form,%
\begin{equation}
q\left( z\right) =-1+\frac{\alpha }{1+\beta a^{n}},  \label{DP}
\end{equation}%
where $\alpha $, $\beta $, and $n>0$ are parameters that control the
behavior of the deceleration parameter. The motivation behind the
parameterization of the deceleration parameter in Eq. (\ref{DP}) comes from
the fact that the deceleration parameter is an important cosmological
parameter that characterizes the dynamics of the Universe. It is defined as
the ratio of the cosmic acceleration to the cosmic expansion rate i.e. $q=-%
\frac{\overset{..}{a}}{aH^{2}}$. In the past, the deceleration parameter was
thought to be a constant value, indicating that the Universe is either
slowing down or maintaining a constant rate of expansion. However,
observational evidence in recent years suggests that the Universe is in fact
accelerating in its expansion, which requires a modification of the
deceleration parameter. To account for this acceleration, various
parameterizations of the deceleration parameter have been proposed in the
literature \cite{Mamon1,Mamon2,Das,Riess1,Akarsu,Santos,Campo,Nair,Gong,Xu,CapozzielloDP}. The parameterization used in this paper is a generalized form
that allows for more flexibility in describing the evolution of the
Universe. The parameter $\alpha $ describes the present value of the
deceleration parameter, while $\beta $ and $n$ determine the functional form
of its evolution with respect to the scale factor $a\left( t\right) $. The
value of the deceleration parameter parameterization depends on the values
of $\alpha $, $\beta $, $n$, and the scale factor $a=\left( 1+z\right)
^{-1}$ at a particular redshift $z$. However, we can give some
general properties about the behavior of $q$ at different redshifts: 

\begin{itemize}
\item At the present time ($z=0$), the current value of the deceleration
parameter $q_{0}$ is given by $q_{0}=-1+\frac{\alpha }{1+\beta }$. This
means that the present-day acceleration or deceleration of the Universe
depends on the value of $\alpha $ and $\beta $. If $\alpha <1+\beta $, the
Universe is currently accelerating, if $\alpha >1+\beta $, the Universe is
currently decelerating, and if $\alpha =1+\beta $, the Universe is currently
coasting.

\item In the past ($z>>1$), the Universe was dominated by matter, and the
deceleration parameter $q$ was positive, meaning that the expansion of the
Universe was slowing down. The exact value of $q$ is $q_{>>1}=\alpha -1$
i.e. depends on the value of $\alpha $.

\item In the far future ($z\rightarrow -1$), if the Universe continues to
expand at an accelerating rate, the deceleration parameter $q$ will approach 
$-1$ asymptotically. It indicates that the Universe is entering a phase of
exponential expansion, known as the De Sitter phase. 
\end{itemize}

The deceleration parameter parametrization used in our model is in agreement
with the thermodynamic constraints on the deceleration parameter presented
in the article \cite{CapozzielloDP}. This indicates that our model can accurately capture the
thermodynamic behavior of the Universe.

The equation below establishes a relationship between the deceleration
parameter and the Hubble parameter,%
\begin{equation}
H\left( z\right) =H_{0}\exp \left( \int_{0}^{z}\frac{1+q\left( z\right) }{%
\left( 1+z\right) }dz\right) .  \label{Hq}
\end{equation}

By substituting Eq. (\ref{DP}) into Eq. (\ref{Hq}), we obtain the expression
for $H\left( z\right) $ as,%
\begin{equation}
H\left( z\right) =H_{0}\left( 1+z\right) ^{\alpha }\left( \frac{1+\beta
\left( 1+z\right) ^{-n}}{1+\beta }\right) ^{\frac{\alpha }{n}},
\label{Hz}
\end{equation}%
where $H_{0}$ represents the current value of the Hubble parameter (at $z=0$). The time derivative of the Hubble parameter can be written as,%
\begin{equation}
\overset{.}{H}=\frac{dH}{dt}=-\left( 1+z\right) H\left( z\right) \frac{%
dH\left( z\right) }{dz}.
\label{dH}
\end{equation}

For the model parameterization, Eq. (\ref{dH}) becomes,
\begin{equation}
\overset{.}{H}=-\frac{\alpha  H_{0}^2 (1+z)^{2 \alpha +n}}{\beta +(1+z)^n}\left(\frac{1+\beta  (1+z)^{-n}}{1+\beta}\right)^{\frac{2 \alpha }{n}}.
\end{equation}

The model parameterization given in Eq. (\ref{Hz}) allows for a flexible and comprehensive study of the cosmological parameter, which is essential for investigating the behavior and evolution of the Universe. To ensure the model's reliability and consistency with the latest observations, the model's parameters ($H_{0}$, $\alpha$, $\beta$, $n$) are constrained using recent observational datasets. By analyzing the observational data, we can investigate the nature and behavior of the Universe, including the rate of expansion, cosmic acceleration, deceleration, and scalar field EoS. This analysis enables us to test the model's validity and establish a robust and reliable parameterization of the deceleration parameter for future cosmic evolution. The results of this study can significantly enhance our understanding of the Universe's evolution and provide crucial insights into the fundamental cosmological questions.

Expressions of all cosmological parameters in terms of redshift can be found in the appendix \ref{app} of this document. 

\section{Constraints from observational data}
\label{sec3}

The Bayesian technique is a widely used method in cosmology to analyze observational data and extract the values of the cosmological parameters. In this study, we use the Bayesian method along with the Markov Chain Monte Carlo (MCMC) technique to obtain the posterior distributions of the parameters $H_{0}$, $\alpha$, $\beta$, and $n$. The MCMC method is used to generate samples from the posterior distribution, and the \textit{emcee} package is employed to perform the MCMC analysis \cite{Mackey/2013}. To complete the simulation, we use a combination of observational datasets, including the Cosmic Chronometer (CC) Sample from Hubble measurements, the Pantheon sample from Supernovae (SNe), and Baryon Acoustic Oscillations (BAO). The use of multiple datasets ensures the robustness and reliability of our results. The objective is to find the best fit of the parameters, and for this purpose, we use the following likelihood function:
\begin{equation}
    \mathcal{L} \propto exp(-\chi^2/2)
\end{equation}
where $\chi
^{2}$ is the chi-squared function. The $\chi
^{2}$ functions corresponding to various datasets are presented below.

\subsection{$CC$ $dataset$}

We adopt a dataset comprising 31 data points acquired through the use of the CC technique. This method enables us to directly extract information about the Hubble function at various redshifts, extending up to $z \leq 2$. The choice to employ CC data is primarily rooted in its reliance on measurements of age differences between two passively evolving galaxies that originated simultaneously but are separated by a small redshift interval. This approach facilitates the calculation of $\Delta z/\Delta t$. It's noteworthy that CC data has demonstrated higher reliability compared to other methods reliant on absolute age determinations for galaxies \cite{Jimenez1}. The CC data points we have employed were gathered from references \cite{Zhang,Jimenez2,Moresco1,Simon,Moresco2,Stern,Moresco3}, and these references are independent of the Cepheid distance scale and any particular cosmological model. However, it is important to acknowledge that they do rely on the modeling of stellar ages, which is based on robust techniques of stellar population synthesis (for more details, see Refs. \cite{Moresco1,Moresco2,Valent,Corredoira1,Corredoira2,Verde} for analyses related to CC systematics). To assess the goodness of fit of our model with the data, we employ the $\chi ^{2}$ function defined as:
\begin{equation}
\chi^{2}_{CC} = \sum_{i=1}^{31} \frac{\left[H( z_{i},H_{0},\alpha,\beta,n)-
H_{obs}(z_{i})\right]^2}{\sigma^2(z_{i})}.
\end{equation}
where $H(z_{i},H_{0},\alpha,\beta,n)$ represents the theoretical value of the Hubble parameter at redshift $z_i$ for a particular set of cosmological parameters $H_{0}$, $\alpha$, $\beta$, and $n$, $H_{obs}(z_i)$ is the measured value of the Hubble parameter at redshift $z_i$, and $\sigma (z_{i})$ is the corresponding uncertainty of $H_{i}$.

\subsection{$Pantheon$ $dataset$}

SNe are a valuable tool for studying the Universe's accelerating expansion and the nature of DE. These SNe are produced when a white dwarf star explodes in a binary system and have a distinctive light curve that makes them useful "standard candles" for estimating cosmic distances. By comparing their observed luminosity to their theoretical intrinsic luminosity, we can estimate their distance and plot the expansion history of the Universe. The Pantheon sample is an essential dataset of SNe that includes 1048 data points covering a broad range of redshifts $0.01\leq z
\leq2.26$. It was constructed from the PanSTARSS1 Medium Deep Survey, SDSS, SNLS, and several low-$z$ and HST samples. The dataset has been extensively calibrated to minimize systematic errors and improve distance estimation accuracy, making it an essential resource for modern cosmology research \cite{Scolnic/2018,Chang/2019}.

The $\chi^2$ function for the SNe dataset is expressed as, 
\begin{equation}
\chi _{Pantheon}^{2}=\sum_{i,j=1}^{1048}\Delta \mu _{i}\left( C_{Pantheon}^{-1}\right)
_{ij}\Delta \mu _{j},
\end{equation}%
where $\Delta \mu_{i}=\mu_{\mathrm{th}}-\mu_{\mathrm{obs}}$ is the difference between the theoretical and observed distance modulus, $\mu = m_{B}-M_{B}$ represents the difference between the apparent magnitude $m_{B}$ and the absolute magnitude $M_{B}$, and $C_{Pantheon}^{-1}$ represents the inverse of the covariance matrix of the Pantheon
sample. In our analysis, we have adopted a prior distribution for the absolute magnitude $M_B$ of SNe, with a reference value of $M_B = -19.244 \pm 0.037$ \cite{Camarena1}. The nuisance parameters in the equation above were estimated using the BEAMS with Bias Corrections (BBC) approach \cite{Kessler/2017}. The theoretical value of the distance modulus is calculated as follows:
\begin{equation}
\mu _{th}(z)=5log_{10}\frac{d_{L}(z)}{1Mpc}+25,
\end{equation}%
where
\begin{equation}
d_{L}(z)=c(1+z)\int_{0}^{z}\frac{dy}{H(y,\alpha,\beta,n,H_{0} )},
\end{equation}%
is the luminosity distance that takes into account the attenuation of light due to the expansion of the Universe, and $c$ is the speed of light.

\subsection{$BAO$ $dataset$}

Moreover, we also use the BAO dataset obtained from various surveys, including 6dFGS, SDSS, and LOWZ samples of BOSS \cite{BAO1,BAO2,BAO3,BAO4,BAO5,BAO6}. The surveys have provided highly accurate measurements of the positions of the BAO peaks in galaxy clustering at different redshifts. The BAO characteristic scale can be determined using the sound horizon $r_s$ at the epoch of photon decoupling with redshift $z_{dec}$, which is given by
\begin{equation}\label{4b}
r_{s}(z_{\ast })=\frac{c}{\sqrt{3}}\int_{0}^{\frac{1}{1+z_{\ast }}}\frac{da}{
a^{2}H(a)\sqrt{1+(3\Omega _{b,0}/4\Omega _{\gamma,0})a}}.
\end{equation}

Here, $\Omega _{b,0}$ and $\Omega _{\gamma,0}$ represent the current density values of baryons and photons, respectively. We use six data points for $d_{A}(z_{\ast })/D_{V}(z_{BAO})$ from the sources mentioned in Refs. \cite{BAO1,BAO2,BAO3,BAO4,BAO5,BAO6}, where $z_{\ast }\approx 1091$ represents the redshift value for photon decoupling. The comoving angular diameter distance at decoupling is denoted by 
\begin{equation}
    d_{A}(z_{\ast })=c\int_{0}^{z}\frac{dz'}{H(z')},
\end{equation}
and the dilation scale is given by 
\begin{equation}
    D_{V}(z)=\left[ \frac{czd_{A}^{2}(z)}{H(z)}\right] ^{1/3}.
\end{equation}

The BAO dataset is evaluated using the chi-square function presented in \cite{BAO6}, which is expressed as
\begin{equation}\label{4e}
\chi _{BAO}^{2}=X^{T}C_{BAO}^{-1}X,
\end{equation}
where 
\begin{equation}
X=\left( 
\begin{array}{c}
\frac{d_{A}(z_{\star })}{D_{V}(0.106)}-30.95 \\ 
\frac{d_{A}(z_{\star })}{D_{V}(0.2)}-17.55 \\ 
\frac{d_{A}(z_{\star })}{D_{V}(0.35)}-10.11 \\ 
\frac{d_{A}(z_{\star })}{D_{V}(0.44)}-8.44 \\ 
\frac{d_{A}(z_{\star })}{D_{V}(0.6)}-6.69 \\ 
\frac{d_{A}(z_{\star })}{D_{V}(0.73)}-5.45%
\end{array}%
\right) \,,
\end{equation}%
and $C_{BAO}^{-1}$ is the inverse of the covariance matrix \cite{BAO6}.

\subsection{$CC+Pantheon+BAO$ $dataset$}

The combination of multiple cosmological probes can help to better constrain the parameters of a cosmological model and improve our understanding of the Universe. In this regard, the total joint $\chi_{joint}^{2}$ function is often used to combine the constraints from different probes. For example, to combine the CC, Pantheon, and BAO samples, the total joint $\chi_{joint}^{2}$ function is expressed as
\begin{equation}
\chi_{joint}^{2}=\chi_{CC}^{2}+\chi_{Pantheon}^{2}+\chi_{BAO}^{2}.
\end{equation}
where $\chi_{CC}^{2}$, $\chi_{Pantheon}^{2}$, and $\chi_{BAO}^{2}$ are the $\chi^{2}$ functions for the CC, Pantheon, and BAO samples, respectively. 

The total joint $\chi_{joint}^{2}$ function is used to derive the best-fit values and confidence intervals for the parameters of the cosmological model by minimizing the function with respect to the model parameters. This allows us to determine the values of the model parameters that provide the best fit to the observed data, and to estimate the uncertainties in these values. The resulting constraints on the parameters can be used to test different cosmological parameters such as the deceleration parameter and EoS parameter, and to investigate the properties of the Universe on large scales. To obtain our results, we utilized an MCMC analysis with 100 walkers and 1000 steps. Tab. \ref{tab} and Fig.
\ref{CC+P+BAO} show the best-fit values on the parameters $H_{0}$, $\alpha$, $\beta$, and $n$ with $1-\sigma $ and $2-\sigma $ likelihood contours. The model parameters for the $CC+Pantheon+BAO$ joint dataset were found to have best-fit ranges of $H_{0}=67.94_{-0.72}^{+0.72}$ $km/s/Mpc$, $\alpha=1.46_{-0.087}^{+0.11}$, $\beta=2.7^{+1.4}_{-1.2}$, and $n=4.1^{+2.4}_{-2.2}$. Notably, the best-fit value for $H_0$ is in agreement with the measurements reported by the Planck 2018 experiment, which reported a value of $H_0=67.4\pm 0.5$ $km/s/Mpc$ using the cosmic microwave background radiation \cite{Planck2020}. This agreement between the two independent measurements of $H_0$ lends support to the validity of the model used in this analysis. In addition, we compare our  model parameterization with the $\Lambda$CDM model by examining the evolution of the Hubble parameter $H(z)$ and distance modulus $\mu(z)$ with the constraint values of the model parameters $H_{0}$, $\alpha$, $\beta$, and $n$ from the $CC+Pantheon+BAO$ joint dataset. The results of this analysis are presented in Figs. \ref{ErrorHubble} and \ref{ErrorSNe}, which clearly demonstrate that our model parameterization fits the observational data remarkably well. Furthermore, we found that our model is in good agreement with the evolution of the $\Lambda$CDM model, indicating that it can be considered a viable alternative to the $\Lambda$CDM model.

\begin{widetext}

\begin{table*}[!htbp]
\begin{center}
%\adjustbox{width=0.5\textwidth}{
\begin{tabular}{l c c c c c c c}
\hline\hline 
$datasets$              & $H_{0}$ ($km/s/Mpc$) & $\alpha$ & $\beta$ & $n$ & $q_{0}$ & $z_{tr}$ & $\omega_{0}$ \\
\hline
$Priors$   & $(60,80)$  & $(0,10)$  & $(0,10)$ & $(0,10)$ & $-$ & $-$ & $-$\\

$CC$ & $67.86_{-0.77}^{+0.75}$  & $1.26^{+0.30}_{-0.27}$  & $2.4^{+1.9}_{-1.7}$ & $5.7^{+3.9}_{-3.7}$ & $-0.63^{+0.08}_{-0.07}$ & $0.48^{+0.24}_{-0.24}$ & $-1.10^{+0.09}_{-0.08}$\\

$BAO$   & $67.87_{-0.83}^{+0.86}$  & $1.45_{-0.12}^{+0.16}$  & $2.9^{+2.1}_{-1.9}$ & $4.8^{+3.9}_{-3.2}$ & $-0.63^{+0.1}_{-0.1}$ & $0.47^{+0.2}_{-0.16}$& $-1.11^{+0.1}_{-0.1}$\\

$Pantheon$   & $67.93_{-0.78}^{+0.78}$  & $1.28_{-0.32}^{+0.33}$  & $2.5^{+1.8}_{-1.6}$ & $5.5^{+4.2}_{-3.9}$ & $-0.63^{+0.07}_{-0.06}$ & $0.49^{+0.27}_{-0.26}$ & $-1.12^{+0.06}_{-0.05}$\\

$Joint$   & $67.94_{-0.72}^{+0.72}$  & $1.46_{-0.087}^{+0.11}$  & $2.7^{+1.4}_{-1.2}$ & $4.1^{+2.4}_{-2.2}$ & $-0.61^{+0.08}_{-0.07}$ & $0.54^{+0.19}_{-0.17}$ & $-1.09^{+0.08}_{-0.07}$\\

\hline\hline
\end{tabular}
%}
\caption{Cosmological parameter constraints from MCMC analysis: CC, Pantheon, and BAO datasets.}
\label{tab}
\end{center}
\end{table*}

\begin{figure}[h]
\centerline{\includegraphics[scale=0.5]{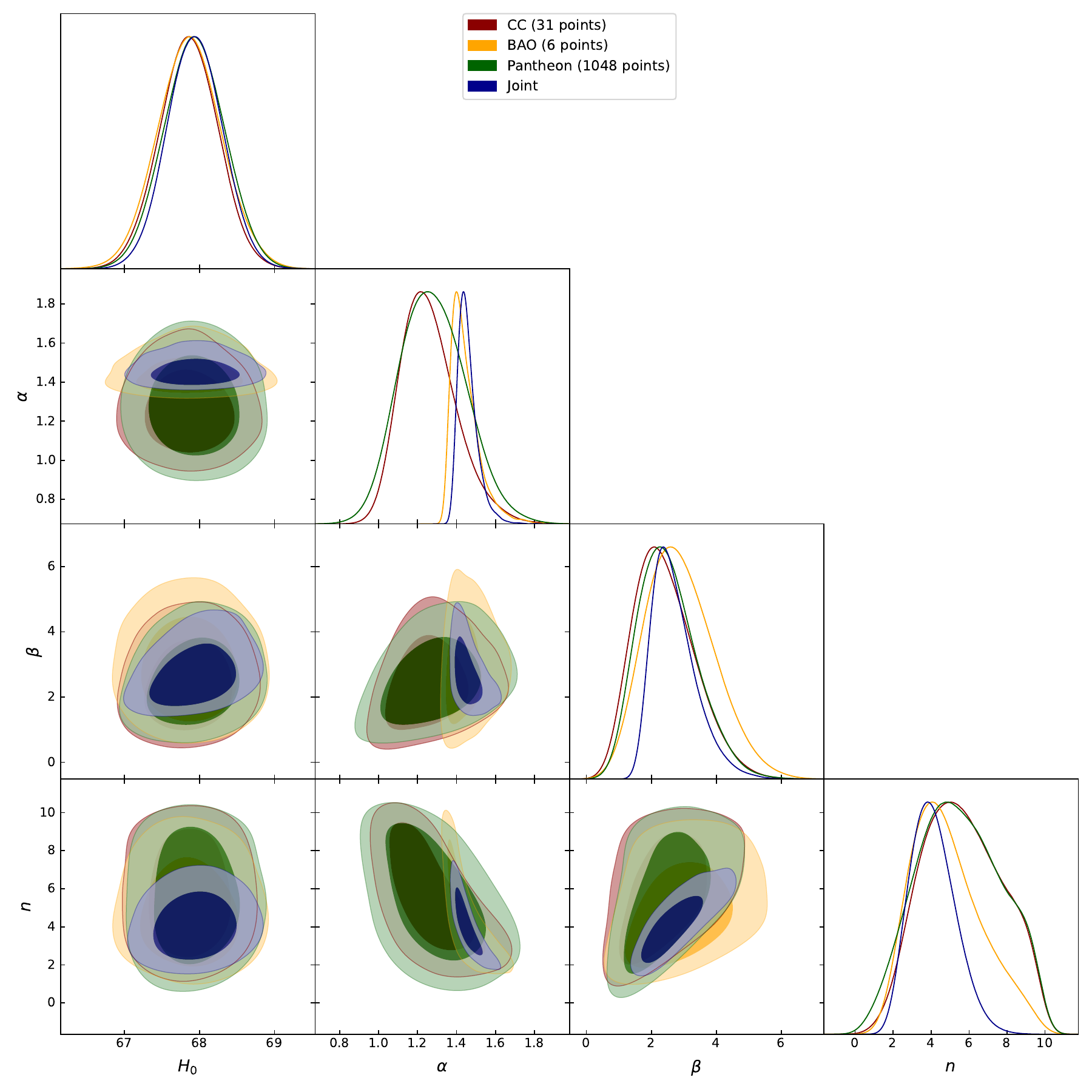}}
\caption{Contour plot of the joint likelihood function for the model parameters $H_{0}$, $\alpha$, $\beta$, and $n$ using $CC$, $Pantheon$, and $BAO$ data with $1-\sigma $ and $2-\sigma $ confidence levels.}
\label{CC+P+BAO}
\end{figure}

\begin{figure}[h]
   \begin{minipage}{1.0\textwidth}
     \centering
     \includegraphics[width=0.9\linewidth]{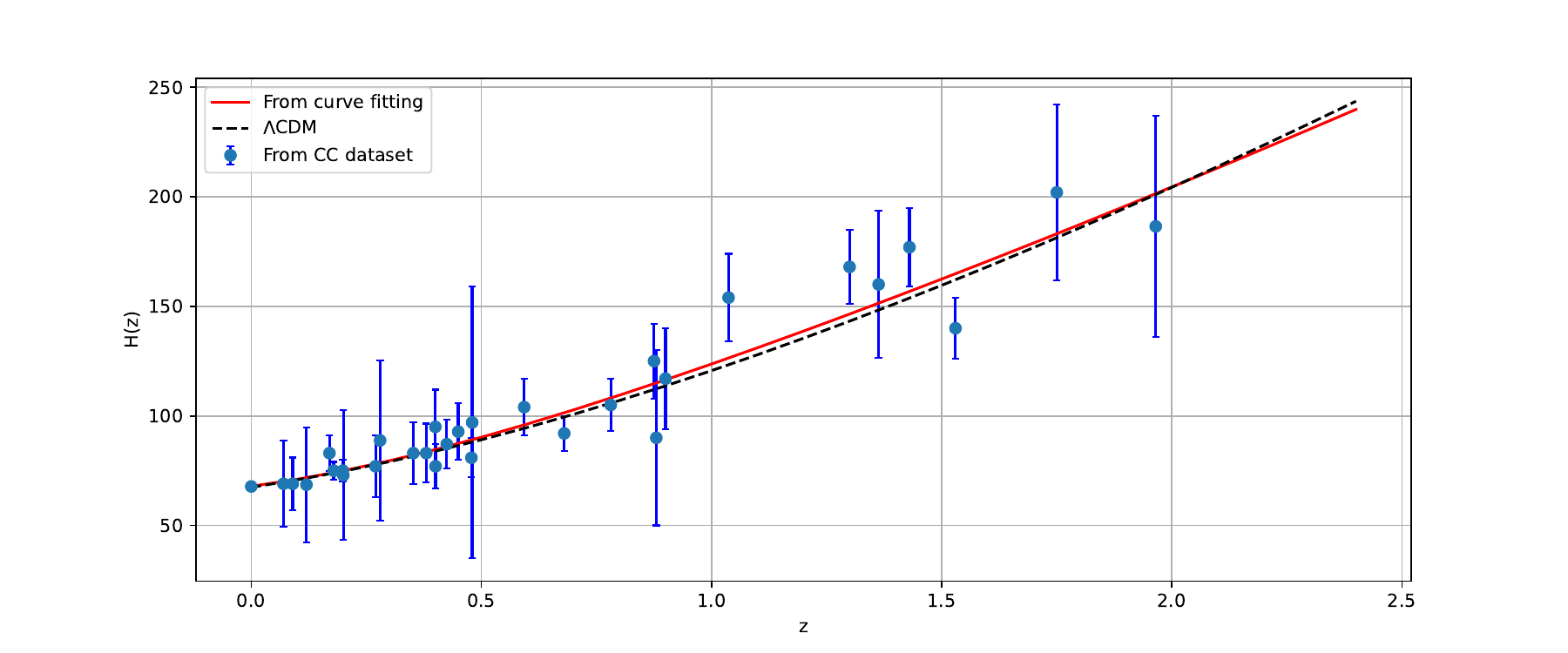}
     \caption{Comparison of the model and $\Lambda$CDM for the Hubble parameter $H(z)$ as a function of redshift $z$. The red line represents the model curve while the black dotted line depicts the $\Lambda$CDM model with $\Omega_{m0}=0.315\pm0.007$. The blue dots with error bars illustrate the 31 CC sample points.}\label{ErrorHubble}
   \end{minipage}\hfill
   \begin{minipage}{1.0\textwidth}
     \centering
     \includegraphics[width=0.9\linewidth]{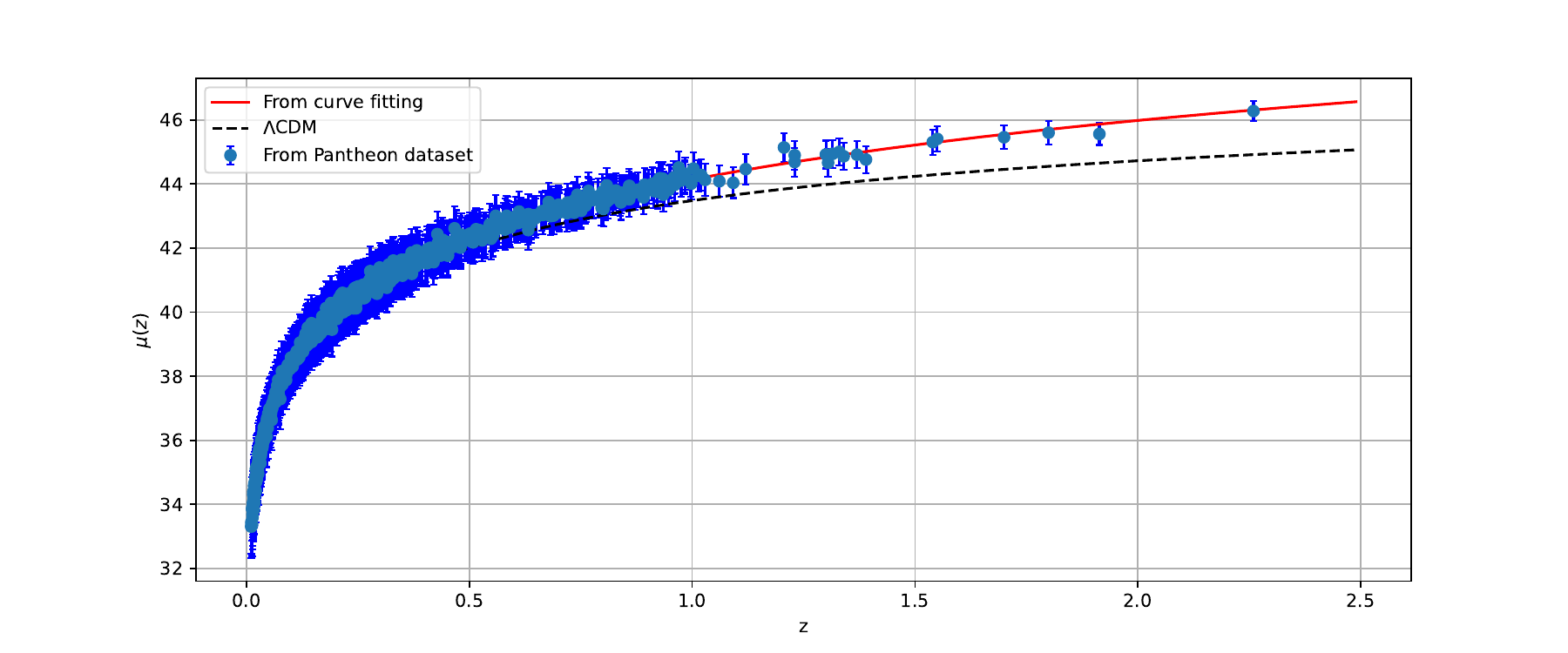}
     \caption{Comparison of the model and $\Lambda$CDM for the distance modulus $\mu(z)$ as a function of redshift $z$. The red line represents the model curve while the black dotted line depicts the $\Lambda$CDM model with $\Omega_{m0}=0.315\pm0.007$. The blue dots with error bars illustrate the 1048 Pantheon sample points.}\label{ErrorSNe}
   \end{minipage}
\end{figure}

\end{widetext}

\section{Analysis Results}
\label{sec4}
The evolution of the cosmological parameters corresponding to the constrained values of the model parameters from the $CC+Pantheon+BAO$ joint dataset are presented below. These parameters include the deceleration parameter, EoS parameter, and density parameter. The obtained values for these parameters are found to be in good agreement with the latest observational and theoretical results.

\begin{widetext}

\begin{figure*}[htb]
\begin{subfigure}{.3\textwidth}
\includegraphics[width=\linewidth]{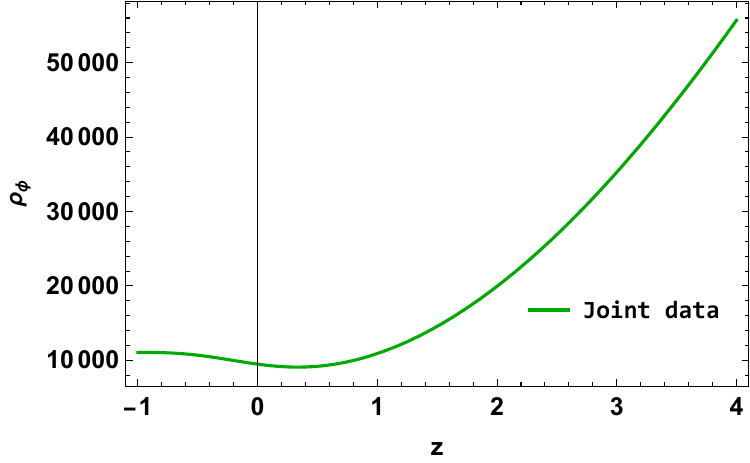}% width
    \caption{DE energy density}
    \label{F_rho_phi}
\end{subfigure}
\hfil
\begin{subfigure}{.3\textwidth}
\includegraphics[width=\linewidth]{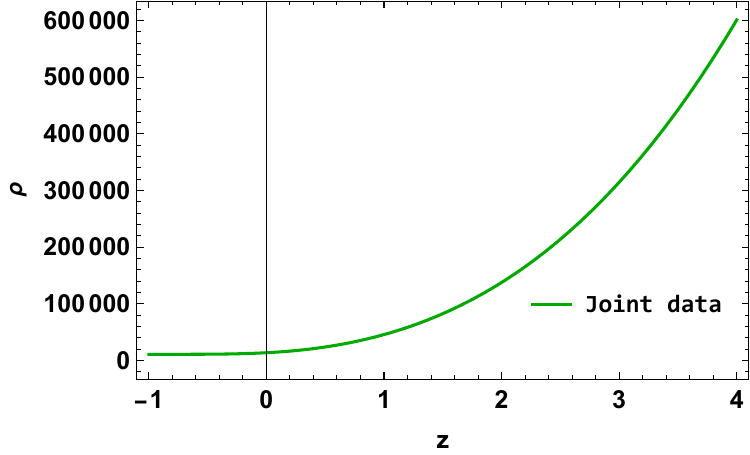}% width
    \caption{Total energy density}
    \label{F_rho}
\end{subfigure}
\hfil
\begin{subfigure}{.3\textwidth}
\includegraphics[width=\linewidth]{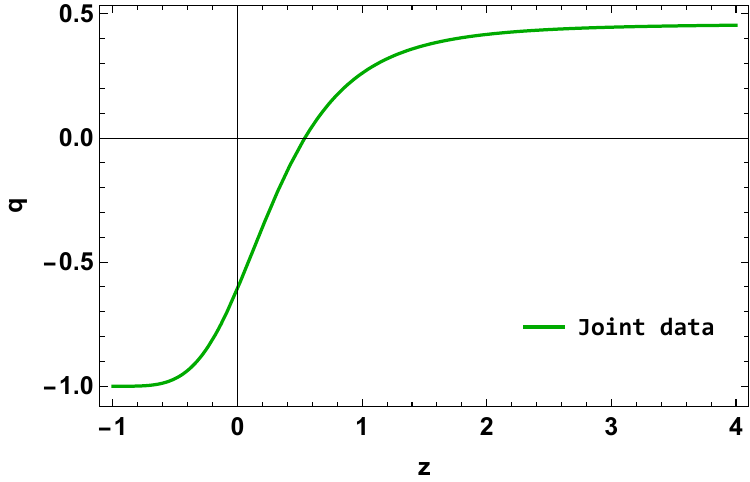}% width
    \caption{Deceleration parameter}
     \label{F_q}
\end{subfigure}

\begin{subfigure}{.3\textwidth}
\includegraphics[width=\linewidth]{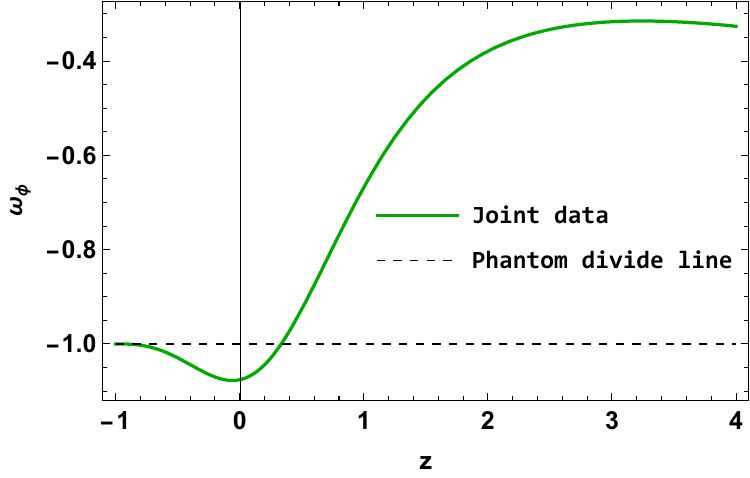}% width
    \caption{DE EoS parameter}
    \label{F_EoS_phi}
\end{subfigure}
\hfil
\begin{subfigure}{.3\textwidth}
\includegraphics[width=\linewidth]{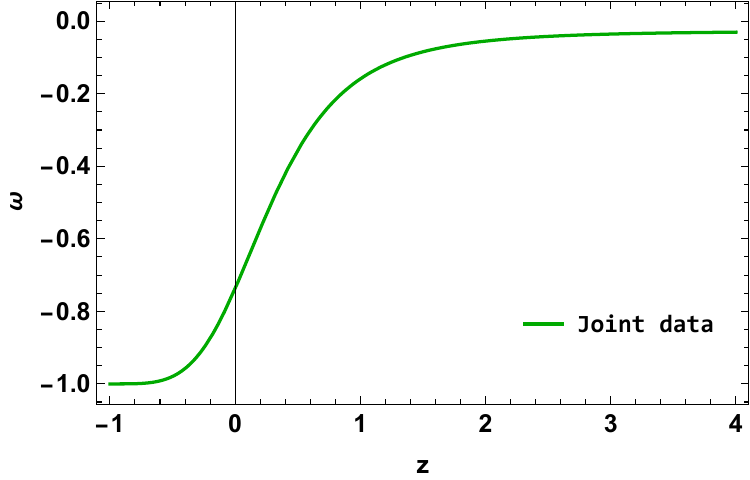}% width
    \caption{Total EoS parameter}
    \label{F_EoS}
\end{subfigure}
\hfil
\begin{subfigure}{.3\textwidth}
\includegraphics[width=\linewidth]{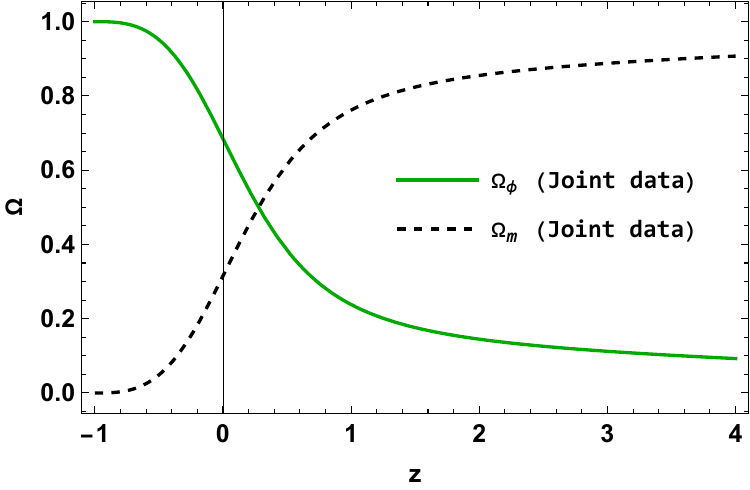}% width
    \caption{Density parameters}
    \label{F_Omega}
\end{subfigure}

\begin{subfigure}{.3\textwidth}
\includegraphics[width=\linewidth]{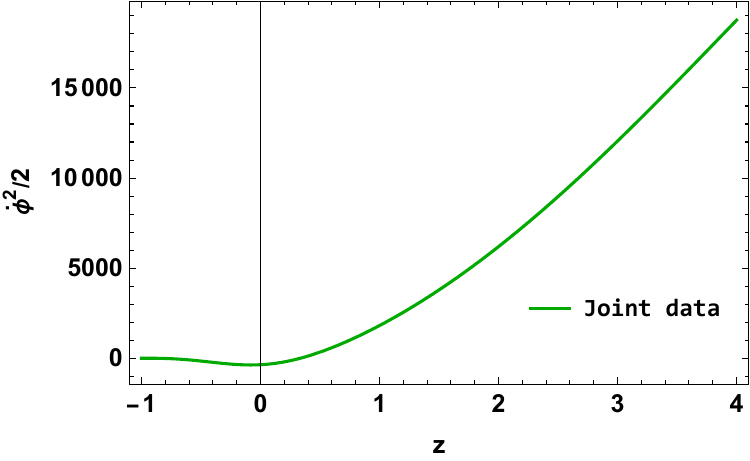}% width
    \caption{Kinetic energy}
    \label{F_phi}
\end{subfigure}
\hfil
\begin{subfigure}{.3\textwidth}
\includegraphics[width=\linewidth]{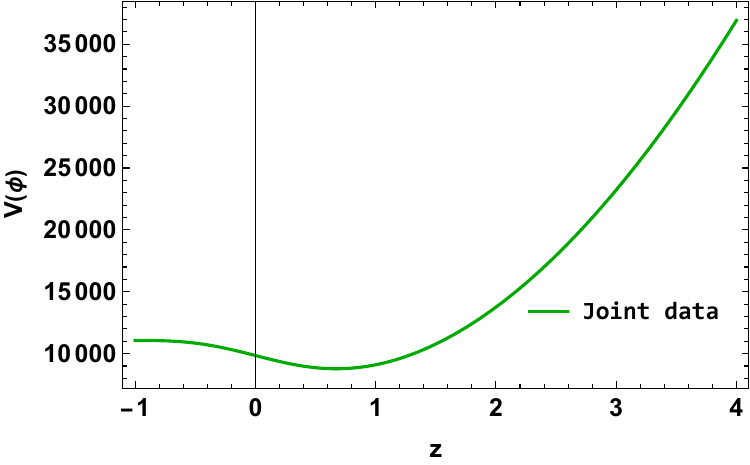}% width
    \caption{Potential energy}
    \label{F_V}
\end{subfigure}
\hfil
\begin{subfigure}{.3\textwidth}
\includegraphics[width=\linewidth]{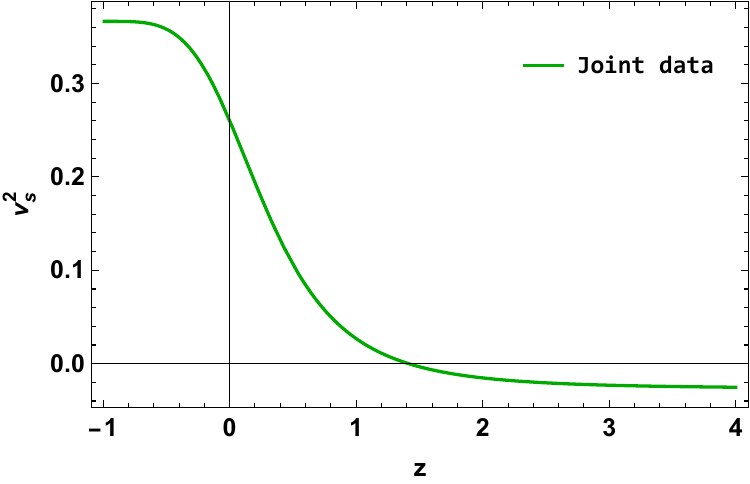}% width
    \caption{Squared speed of the sound}
    \label{F_vs}
\end{subfigure}

\caption{Variation of the cosmological parameters as a function of the redshift $z$ for the model parameters constrained by the $CC+Pantheon+BAO$ joint dataset.}
    \end{figure*}
    
\end{widetext}

Figs. \ref{F_rho_phi} and \ref{F_rho} depict the positive nature of both the DE energy density and the total energy density. Moreover, it is evident that as the Universe evolves, these energy densities gradually decrease. This is consistent with our understanding of the Universe's expansion and the fact that the total matter content of the Universe becomes diluted as the Universe expands. It can be observed that the DE energy density converges to a diminishing value in the far future. This is evident from the behavior of the curve as the redshift ($z$) approaches -1. Furthermore, the evolution of the deceleration parameter presented in Fig. \ref{F_q} shows that the Universe has experienced a transition i.e. at $q=0$ from a decelerated period ($q>0$) to an accelerated period ($q<0$) in the recent past, which is consistent with the current standard model of cosmology. This transition is thought to be driven by DE, which is a mysterious component that dominates the energy density of the Universe and causes its accelerated expansion. The behavior of the deceleration parameter provides important clues about the nature of DE, and understanding it is one of the central goals of modern cosmology. For the model parameters constrained by the $CC+Pantheon+BAO$ joint dataset, the transition redshift is estimated to be $z_{tr}=0.54^{+0.19}_{-0.17}$ \cite{Mamon3,Dinda,Cunha}, and the present value of the deceleration parameter is $q_{0}=-0.61^{+0.08}_{-0.07}$ \cite{Almada,Camarena}. 

The EoS parameter is a key factor in describing the evolution of the Universe dominated by different forms of energy. The present state of the Universe can be predicted by three possible phases, namely the cosmological constant phase ($\omega=-1$), quintessence phase ($-1<\omega<-1/3$), or phantom phase ($\omega<-1$). Moreover, we present the scalar field EoS parameter and total EoS parameter in Figs. \ref{F_EoS_phi} and \ref{F_EoS}, respectively. It is clear that the total EoS parameter initiates in a region dominated by matter and subsequently traverses the quintessence phase before eventually converging to a constant value in the $\Lambda$CDM region. This progression suggests a dynamic evolution of the Universe's energy content, transitioning from a matter-dominated era to a quintessence-dominated phase and ultimately settling into the equilibrium described by the $\Lambda$CDM model. However, the scalar field EoS parameter exhibits a phantom-like behavior. For the model parameters constrained by the $CC+Pantheon+BAO$ joint dataset, the present value of the scalar field EoS parameter is $\omega_{0}=-1.09^{+0.08}_{-0.07}$, 
which is in agreement with previous studies \cite{Novosyadlyj,Kumar,Gong}. 

The behavior of the Universe in its early and late stages can be understood from Fig. \ref{F_Omega}. Initially, during the early period, matter dominates the Universe while the density parameter of the scalar field (or DE) $\Omega_{\phi}$ remains small. As the Universe expands, the matter density parameter $\Omega_{m}$ gradually decreases due to the expansion, while the scalar field density parameter becomes increasingly dominant. This eventually results in the acceleration of the Universe's expansion, which is consistent with the observed late-time cosmic acceleration of the Universe in various cosmological surveys.

The scalar field, known as DE, is an enigmatic form of energy that exists throughout the Universe and is believed to be responsible for the accelerating expansion of the Universe. Figs. \ref{F_phi} and \ref{F_V} illustrate the progression of the scalar field's kinetic and potential energy. As time elapses, the scalar field transitions from a state of high energy to one of lower energy. This transformation is evident in the diminishing values of both kinetic and potential energy, as they shift from high positive values to lower positive values.

\section{Stability analysis}
\label{sec5}

Our current focus is on examining the stability of the obtained model when subjected to perturbations. Initially, we investigated the dynamic behavior of the model. Now, we aim to determine whether small disturbances in the background result in the amplification or reduction of the perturbations. According to perturbation theory, the classical stability or instability of the model is determined by the sign of the squared sound speed ($v_{s}^2$). A positive value of $v_{s}^2$ indicates stability, as perturbations propagate within the background. Conversely, a negative value of $v_{s}^2$ signifies instability, as even small perturbations grow exponentially within the background.

To determine the value of the squared sound speed ($v_{s}^2$), we can use the following expression \cite{Kim}:
\begin{equation}
    v_{s}^2=\frac{dp}{d\rho}.
\end{equation}

From Fig. \ref{F_vs}, it is evident that the squared speed of sound is negative in the early stages of the Universe (i.e. at $z>>0$). This indicates an unstable condition within the model during this period. However, for the present and future (i.e. at $z=0$ and $z<0$), the squared speed of sound becomes positive, implying stability within the model. This transition from negative to positive values signifies a significant shift in the dynamics of the Universe, ensuring a stable behavior in the model as time progresses.

\section{Conclusion}
\label{sec6}
In this work, we have explored a cosmological model parametrized by a generalized variable deceleration parameter. By analyzing the model with the constraints from the latest observational datasets, including CC, Pantheon, and BAO, we have found that the model is consistent with the data within the $1-\sigma$ and $2-\sigma$ confidence levels. The model parametrization utilized in this study aligns with the thermodynamic constraints imposed on the deceleration parameter, as discussed in the literature \cite{CapozzielloDP}. This finding implies that our model can successfully capture the thermodynamic behavior of the Universe.

Our results show that both the
DE energy density and the total energy density are positive and decrease with time, while the deceleration parameter indicates that the Universe has transitioned from a decelerated period to an accelerated period in the recent past. The EoS parameter of the scalar field exhibits phantom-like behavior and approaches the cosmological constant at lower redshifts. Further, both kinetic and potential energy transition from a state of high energy to one of lower energy with time. For the $CC+Pantheon+BAO$ joint dataset, the present value of the Hubble parameter is $H_{0}=67.94_{-0.72}^{+0.72}$ $km/s/Mpc$, which is consistent with the latest Planck results \cite{Planck2020}. Also, based on the constraints obtained from the joint dataset, we have found that the transition redshift is estimated to be $z_{tr}=0.54^{+0.19}_{-0.17}$ \cite{Mamon3,Dinda,Cunha}, indicating a transition from a decelerating phase to an accelerating phase. Furthermore, the present value of the deceleration parameter is determined to be $q_{0}=-0.61^{+0.08}_{-0.07}$ \cite{Almada,Camarena}, implying that the Universe is currently experiencing accelerated expansion. These values are consistent with observational data from various surveys, such as Pantheon \cite{Scolnic/2018}. Additionally, the EoS parameter for the scalar field is $\omega_{0}=-1.09^{+0.08}_{-0.07}$ \cite{Novosyadlyj,Kumar,Gong}. So, the values obtained in our model are consistent with other studies and observational data, indicating that the model can accurately describe the thermodynamic behavior of the Universe.

Finally, in order to assess the stability of the model, we conducted a classical stability analysis by examining the squared sound speed. Our findings indicate that the model remains stable against small perturbations both in the present and future, as illustrated in Fig. \ref{F_vs}. This result is consistent with a previous study conducted by \cite{Huang}, further reinforcing the stability of our model.

\section*{Acknowledgments}
The authors extend their appreciation to the Deanship of Scientific Research, Imam Mohammad Ibn Saud Islamic University (IMSIU), Saudi Arabia, for funding this research work through Grant No. (221412042).

\textbf{Data availability} There are no new data associated with this
article.

\begin{widetext}
\section*{Appendix}
\label{app}

DE EoS parameter:
\begin{equation*}
    \omega _{\phi}\left( z\right) =\frac{(2 \alpha -3) (1+z)^n-3 \beta }{3 \left(\beta +(1+z)^n\right) \left(1-\Omega_{m0} (1+z)^{3-2 \alpha } \left(\frac{\beta  (1+z)^{-n}+1}{\beta +1}\right)^{-\frac{2 \alpha }{n}}\right)}.
\end{equation*}

Total EoS parameter:
\begin{equation*}
   \omega(z) =-1+\frac{2 \alpha  (1+z)^n}{3 \left(\beta +(1+z)^n\right)}.
\end{equation*}

Matter density parameter:
\begin{equation*}
   \Omega_{m}(z) =\Omega_{m0} (1+z)^{3-2 \alpha } \left(\frac{\beta  (1+z)^{-n}+1}{\beta +1}\right)^{-\frac{2 \alpha }{n}}.
\end{equation*}

Kinetic energy:
\begin{equation*}
   \frac{1}{2}\overset{.}{\phi }^{2} =\frac{\alpha  H_{0}^2 (1+z)^{2 \alpha +n} \left(\frac{\beta  (1+z)^{-n}+1}{\beta +1}\right)^{\frac{2 \alpha }{n}}}{\beta +(1+z)^n}-\frac{3}{2} H_{0}^2 \Omega_{m0} (1+z)^3.
\end{equation*}

Potential energy:
\begin{equation*}
   V(\phi)=H_{0}^2 (1+z)^{2 \alpha } \left(\frac{\beta  (1+z)^{-n}+1}{\beta +1}\right)^{\frac{2 \alpha }{n}} \left(3-\frac{\alpha }{\beta  \left(\frac{1}{1+z}\right)^n+1}\right)-\frac{3}{2} H_{0}^2 \Omega_{m0} (1+z)^3.
\end{equation*}

Squared speed of the sound:
\begin{equation*}
   v_{s}^2(z)=\frac{\beta  (n-3)+(2 \alpha -3) (1+z)^n}{3 \left(\beta +(1+z)^n\right)}.
\end{equation*}

\end{widetext}

\end{document}